\documentclass[reprint,amsmath,amssymb,aps,showkeys]{revtex4-1}
\usepackage{subfigure}
\usepackage{graphicx}
\usepackage{multirow}
\usepackage{bm}
\usepackage{lineno}

\begin{document}
\preprint{APS/123-QED}

\title{On Deriving  Probabilistic Models for Adsorption Energy on Transition Metals using Multi-level \textit{ab initio} and Experimental Data}
\author{Huijie Tian}
\affiliation{Department of Chemical and Biomolecular Engineering, Lehigh University}
\author{Srinivas Rangarajan}
\affiliation{Department of Chemical and Biomolecular Engineering, Lehigh University}
\email{srr516@lehigh.edu}

\begin{abstract}
In this paper, we apply multi-task Gaussian Process (MT-GP) to show that the adsorption energy of small adsorbates on transition metal surfaces can be modeled to a high level of fidelity using data from multiple sources, taking advantage of the relatively abundant ''low fidelity" data (such as from density functional theory computations) and small amounts of ''high fidelity" computational (e.g. using the random phase approximation) or experimental data. To fully explore the performance of MT-GP, we perform two case studies - one using purely computational datasets and the other using a combination of experimental and computational datasets. In both cases, the performance of MT-GPs is significantly better than single-task models built on a single data source. This method can be used to learn improved models from fused datasets, and thereby build accurate models under tight computational and experimental budget.
\end{abstract}

\maketitle

\section{Introduction}
The adsorption (or binding) energy of molecules and molecular fragments is a fundamental property of interest in heterogeneous catalysis and surface science.\cite{norskov2009towards, norskov2011density} Quantifying the energetics of closed and open shell species and surface reactions on a catalyst enables (1) elucidating the underlying reaction mechanism by building predictive kinetic models\cite{gokhale2004molecular} and (2) analyzing the reactivity and selectivity trends across a space of potential alternative catalytic formulations.\cite{norskov2009towards, greeley2004alloy, norskov2002universality} The binding energies can be obtained from experimental measurements\cite{silbaugh2016energies, campbell2013enthalpies} or quantum chemistry calculations\cite{norskov2011density, gokhale2004molecular}. Several experimental measurement techniques, such as Temperature Programmed Desorption (TPD) and Single Crystal Adsorption Calorimetry (SCAC), provide accurate measurements of the adsorption energy on transition metals and metal oxides with errors from 0.02 to 0.1 eV.\cite{silbaugh2016energies, campbell2013enthalpies} On the other hand, quantum chemistry, especially density functional theory (DFT), is commonly employed in catalysis research to calculate the binding energies of adsorbates. However, DFT functionals do not offer the much coveted chemical accuracy.\cite{wellendorff2015benchmark, gautier2015molecular} Recently, a benchmarking study evaluated the accuracy of six commonly used functionals on a dataset of thirty nine surface species on single-crystal transition metal facet which are accurately measured experimentally.\cite{wellendorff2015benchmark} The mean error of Generalized Gradient Approximation (GGA) functionals (i.e. PW91, PBE, RPBE, BEEF-vdW, etc.) ranged from 0.2 to 0.4 eV. More advanced theories, such as the Random phase approximation (RPA), outperform commonly used functional for adsorption energies.\cite{schmidt2018benchmark, yan2018hydrogen, tameh2018accuracy} For instance, Schmidt et al. \cite{schmidt2018benchmark} compared the binding energy of eight adsorbates on twenty five different metals at full monolayer coverage using RPA and a number of common functionals (PBE, RPBE, BEEF-vdW). They showed that the mean absolute deviations between these functionals and RPA is about 0.2 eV. 

However, the computational cost of applying RPA is significantly higher than typical GGA-based functionals, precluding its use to explore medium to large reaction networks. Further, while relatively scalable, even GGA functionals are too expensive to use while exhaustively exploring the potential energy surface of a complex reaction network comprising of thousands of entities.\cite{ulissi2017address, simm2018error}

Statistical and machine learning can help address the challenge of developing "cheap" surrogates that offer a high degree of accuracy relative to high levels of theory or experiments.\cite{ulissi2017address, simm2018error, rangarajan2014automated, rangarajan2014identification, vinu2012unraveling} Several machine learning strategies have been proposed for modeling the adsorption energy. Linear scaling relations, the simplest class of data-driven models\cite{abild2007scaling, bronsted1928acid, evans1938inertia, greeley2016theoretical, calle2015introducing}, have been proposed and widely used for catalyst design\cite{norskov2009towards} and reaction network exploration\cite{ulissi2017address, rangarajan2014identification}. Group additivity model and feature selection algorithm have been used for modeling the adsorption energy on metals.\cite{gu2018thermochemistry, rangarajan2014automated} Gaussian Process has been applied in constructing surface phase diagram for metal oxide and sulfide applied in electrochemical catalysis.\cite{ulissi2016automated} More advanced machine learning tools, such as artificial neural networks, has been used to model the CO binding energy on transition metal alloy, then screening catalyst for CO$_2$ electroreduction.\cite{ma2015machine} 

A data-driven surrogate model trained on computational datasets inherits deficiencies from the original level of theory. To improve the predictive accuracy of the surrogate, a high level of theory, such as RPA, or experimental data is necessary. However, the computational cost increases significantly as one chooses progressively higher levels of theory, as illustrated in \textit{Jacob's ladder of density functional approximation}\cite{perdew2001jacob}, which can be a huge concern when building the training set for a statistical model. 

Multi-fidelity modeling is well-suited to address the problem of computational cost. Specifically, it enables building a flexible statistical model with large amounts of less accurate data and relatively little data from high level of theory and experimental data such that the final model has a higher accuracy than any model trained on any single set of data (especially the limited high fidelity dataset). There are multiple examples of multi-fidelity modeling in the literature. Lilienfeld and his collaborators developed a statistical model, called $\Delta$-Machine Learning \cite{ramakrishnan2015big}, wherein they built a kernel ridge regression model on the difference of the atomization energy from two levels of theory. Further, they extended the framework not only for the levels of theory (HF, MF2, CCSD(T), etc.), but also for the choice the basis set (6-31(g), cc-pvdz, etc.).\cite{zaspel2018boosting} Pilania et al. have taken multi-task Gaussian Process approach to model the bandgap energy from two levels of theory, viz. PBE and HSE06, and they showed the improvement of model prediction with increasing number of PBE data.\cite{pilania2017multi} In the context of catalysis, Xu and his collaborators notice that GGA functional will tend to overestimate the chemisorbed species and underestimate the physisorbed species, while van der waals correction functional have the opposite trend. They propose a simple two-level theory based on reweighting the binding energy estimates of two functionals (RPBE and optB-vdW).\cite{hensley2017dft}

In this work, we adopt Gaussian Process (GP) and its multi-task version (MT-GP) to obtain an adsorption energy model using data from multiple sources. Multi-task Gaussian Processes offer more flexibility than approaches considered so far because it allows us to fuse multiple datasets (each at a different level of fidelity)\cite{lee2018linking}. Moreover, it enables modeling datasets with different size, since it learns the correlation between the different sets. We perform two case studies to illustrate the robustness of this approach. In the first case study, we investigate the performance of the GP model on purely computational datasets comprising of binding energies at multiple levels of theory, and explore the model performance with respect to the choice of low level of theory. In the second case, we collect  adsorption energies reported in the literature (PW91 and RPA calculations, and experimental data), to develop a MT-GP model on the fused data. We show that after incorporating low level of theory, the performance of multi-task Gaussian Process significantly improves over a GP model built only on data of single fidelity (single-task GP or ST-GP). This method provides a general framework for fusing multi-level computational and experimental data.

\section{Computational Method}

\subsection{Gaussian Process}
Gaussian Process (GP) is a statistical supervised learning algorithm used for classification and regression, which models the governing function of the data with a stochastic process.\cite{rasmussen2006gaussian} Guassian Process Regression (GPR) has been applied in a number of related problems. It has been used for (1) calibrating the systematic error of computational model \cite{kennedy2001bayesian, campbell2006statistical, pernot2017critical}, such as electrochemical thermal model\cite{tagade2016bayesian}, (2) fitting of potential energy surfaces\cite{kolb2017representing, alborzpour2016efficient, qu2018assessing, alborzpour2016efficient}, (3) exploring  reaction networks\cite{ulissi2017address, simm2018error}, and (4) accelerating structure optimization and transition state search\cite{denzel2018gaussian, koistinen2017nudged, schmitz2018gaussian, torres2018low}, to name a few. A non-probabilistic analog of GP, viz., kernel-ridge regression, has been used for predicting molecular properties in the context of material screening\cite{rupp2012fast, faber2017prediction, ramakrishnan2015big}. 

In contrast to parametric models which have a fixed functional form, GP is a non-parametric model, extrapolating the training data to make prediction. GP defines a multivariate Gaussian distribution over the underlying function, which is governed by the mean and covariance function. Usually, the mean function is assumed to be zeros. The covariance function estimates the correlation and similarity between the data point. Supposing there are $n$ data points, with inputs $\mathbf{X}=(X_1,X_2,...,X_n)$, and output $\mathbf{y}=(y_1,y_2,...,y_n)$, and given a covariance function $k=k(x,x')$, the prediction of given data point $X^*$, $y^*$ follows a joint Gaussian distribution such as,

\begin{equation}
\label{eqn:stgp_pred}
y^*|\mathbf{y}=\mathcal{N}(K_{*}K^{-1}\mathbf{y}, K_{**}-K_*K^{-1}K_*^\mathrm{T})
\end{equation}
where $K$ is the kernel matrix, obtained from applying the kernel function on each pair of the inputs, $K_*$ is a vector which is from applying the kernel function on the new data point and all training data, and $K_{**}$ is a value, applying the kernel function on the new data point itself. We could see that the mean of the prediction is a weighted sum of the training data, and the weights depend on the similarity of data points, calculated by the kernel matrix. Further, the variance of the unseen (new) data depends on the kernel matrix, and is independent of the output of the training data ($\mathbf{y}$, $y^*$).

Covariance function, also known as kernel function, is the most essential part of a GP model, which defines the correlation between the data points. The choice of covariance function represents the prior assumption on the underlying function, such as the smoothness, periodicity, etc. The selection of kernel function is often done by trial-and-error, while recent research has been focusing on automatically constructing a kernel function by searching in a function space\cite{duvenaud2013structure}. For this study, we compared the performance of different kernels, which all presented similar accuracy and finally settled on the Squared Exponential (SE) kernel as the covariance function due to its smoothness and differentiability. This function is expressed as 

\begin{equation}
\label{eqn:stgp_rbf}
k(x, x') = \sigma_f^2 exp(-\frac{\| x - x'\|_2^2}{l^2})+\sigma_{noise}^2
\end{equation}
where $l$ is the characteristic lengthscale of the input, $\sigma_f$ and $\sigma_{noise}$ are the variance of the kernel and white noise. These are the hyperparameters that need to be optimized. Assuming the likelihood function is a Gaussian distribution, the analytical expression of the marginal likelihood is available, such as, 

\begin{equation}
\label{eqn:stgp_mll}
\mathrm{log} P(\mathbf{y}|\mathbf{X},\mathbf{\theta}) = -\frac{1}{2}y^\mathrm{T}K^{-1}y-\frac{1}{2}\mathrm{log}|K|-\frac{n}{2}\mathrm{log}2\pi
\end{equation}
where $\mathbf{\theta}$, the vector of hyperparameters, is calculated via minimizing the negative marginal likelihood. Since the function is not convex, using single starting point for optimization does not guarantee global minimum. Usually, a multi-start sampling of the initial values of these hyperparameters are required to efficiently search the parameter space.

There are several advantages of applying GP on the adsorption energy dataset. First, GP is a probabilistic model, which is able to make a point estimate and further provide the variance of the unseen data point. Moreover, the predictive variance can be easily applied in Bayesian optimization\cite{snoek2012practical} and active learning framework\cite{kapoor2007active, tang2018prediction}. Second, while GP scales as $\mathcal{O}(n^3)$ with respect to the size of the training set, $n$, for training, rendering it inefficient for large datasets, it is well suited for adsorption energy dataset, where the size of train set is small to medium, usually less than one thousand. Third, GP is flexible in modeling highly nonlinear functions. Fourth, it is easier to train compared to deep neural networks. We refer to models described herein as a ``single task" GP (or ST-GP).

\subsection{Multi-task Gaussian Process}

In this section, we will go through a brief overview of the multi-task Gaussian Process (MT-GP)\cite{bonilla2008multi}, also known as coregionalized Gaussian Process\cite{alvarez2012kernels}. Rather than handling scalar output in ST-GP, MT-GP is dealing with vector-valued output. In the machine learning community, this modeling technique is considered as \textit{multitask} learning, where several related tasks could transfer information between each other to improve the accuracy of all tasks.\cite{pan2010survey} Let us suppose that we have $m$ related tasks, with input-output pairs $(\mathbf{X}^1, \mathbf{y}^1), (\mathbf{X}^2, \mathbf{y}^2),...,(\mathbf{X}^m, \mathbf{y}^m)$, with each task referring to a separate level of theory. For simplicity, we assume the number of data points in each task is the same, viz. $n$, and the vector form of the data is 

\begin{equation}
\label{eqn:mtgp_xy}
\mathbf{X} = \begin{pmatrix}
                \mathbf{X}^1 \\
                \mathbf{X}^2 \\
                \vdots \\
                \mathbf{X}^m
            \end{pmatrix},\quad
\mathbf{y} = \begin{pmatrix}
                \mathbf{y}^1 \\
                \mathbf{y}^2 \\
                \vdots \\
                \mathbf{y}^m
            \end{pmatrix}
\end{equation}

In the multiple output case, the Gaussian process is a collection of different random processes evaluated at each input set, and the distribution of the underlying vector function is given as,
\begin{equation}
\label{eqn:mtgp_intro}
    \mathbf{f}(\mathbf{X})\sim \mathcal{N}(0,\mathbf{K}(\mathbf{X}, \mathbf{X}))
\end{equation}
For a ST-GP, the covariance matrix expresses the correlation, and determines the weights on each data for prediction. In its multi-task version, the composite covariance matrix learns the correlation among all the data points in all the task simultaneously, every data points in all tasks will contribute to the prediction. For simplicity, we apply the intrinsic coregionalization model (ICM)\cite{alvarez2012kernels} wherein the covariance function is expressed as the Kronecker product between the inner-task covariance function and inter-task covariance function, such as

\begin{equation}
\label{eqn:mtgp_K}
\begin{split}
    \mathbf{K}(\mathbf{X},\mathbf{X}) &= \mathbf{B}\otimes k(\mathbf{X},\mathbf{X})\\
    &= 
    \begin{bmatrix}
    b_{11}k(\mathbf{X},\mathbf{X}) & \dots  & b_{1m}k(\mathbf{X},\mathbf{X})\\
    \vdots                         & \ddots & \vdots \\
    b_{m1}k(\mathbf{X},\mathbf{X}) & \dots  & b_{mm}k(\mathbf{X},\mathbf{X})
\end{bmatrix}
\end{split}
\end{equation}
where $\mathbf{B}$ is the coregionalization matrix, and the off-diagonal of $\mathbf{B}$ encodes the correlation between each tasks. Under this setting, the composite covariance matrix is expressed as the product of task similarity and data similarity, not strictly speaking. For example, if two tasks are similar to each other, the prediction for one task will have large weights on the data point in the other task. On the other hand, if the tasks are not similar at all, the off-diagonal block of $\mathbf{K}$ will be zero matrix, and the MT-GP can be expressed as $m$ independent single task Gaussian Processes. In ICM, the characteristic length-scale of the inner task covariance function is assumed to be the same within each task, the scale of the covariance function is controlled by the coregionalization matrix. In this study, we adopt ICM because of its simplicity but note that more general frameworks such as the linear model of coregionalization (LMC) exist to model the complex correlation structure.

\subsection{Datasets}
In this study, the adsorption energies are obtained from three sources: 

(1) \textbf{FC}: Two hundred (200) binding energies comprising of eight adsorbates (O, H, N, CO, OH, NO, CH, N$_2$) on twenty five (25) transition metals FCC (111) facet with monolayer surface coverage (full coverage, or FC), and calculated by several levels of theory. Each set is referred to as FC-RPA, FC-PBE, FC-BEEF.\cite{schmidt2018benchmark}

(2) \textbf{PW91}: Three hundred and seventy seven (377) adsorption energies calculated using the PW91 functional reported by the Mavrikakis group.\cite{PW91_818, PW91_831, PW91_834, PW91_838, PW91_833, PW91_836, PW91_841, PW91_842, PW91_843, PW91_828, PW91_825, PW91_830, PW91_829, PW91_839, PW91_821, PW91_832, PW91_840, PW91_835, PW91_823, PW91_826, PW91_822, PW91_824, PW91_827, PW91_820, PW91_819} The data were collected from fourteen articles on: (i) benchmarking adsorption energy of small adsorbates on a number of transition metal facets, (ii) heterogeneous catalytic reaction trends and (iii) microkinetic modeling studies. The list of species includes single atoms such as C, H, O, N, S, open-shell molecules like OH, COOH, HCOO, which are important intermediates in catalytic reactions, and closed-shell molecules such as CO, CO2, H2O, whose adsorption energies are experimentally measurable. The binding energy ranges from -8 eV with strongly adsorbed species such as nitrogen and carbon atoms, to ~0 eV for weakly adsorbed species including CO$_2$, H$_2$O. This set includes 17 transition metals, and the adsorbed facet includes the closed-packed facets (111), (100), (0001) and a step edge (211). To ensure consistency of the data, we choose the species calculated using the specific setting: (i) the software used is DACAPO, a planewave energy code, (ii) the PW91 functional is used with the plane wave cutoff of 25 Ry, (iii) the ionic cores were represented by Vanderbilt ultrasoft pseudopotentials, (iv) a Fermi smearing was used with the setting $k_BT=0.1$ eV,and (v) the first Brillouin zone is sampled by Chadi–Cohen k-points an, while the number of k-points may vary from one metal to the other, convergence checks ensured the binding energies converged to tens of meV. \textbf{PW91(e)} dataset denotes an expanded dataset comprising of 508 surface species, including 131 additional surface species for which Brillouin zone was sampled with Monkhorst-Pack k points, mainly used for calculating the species on (100) and (211) facet.

(3) \textbf{Exp}: Sixty one (61) surface species experimentally measured or re-evaluated by the Campbell group.\cite{EXP_846, EXP_849, EXP_845, EXP_854, EXP_853, EXP_855, EXP_857, EXP_856, EXP_844, EXP_852, EXP_850, EXP_851, EXP_847, EXP_848} Most of the adsorbates are closed-shell molecules such as CO, H2O, etc., directly measured by TPD or SCAC; rest of the species are single atoms and open-shell molecules such as O, H, CH$_3$, HCOO, OH, etc., which are measured by dissociative adsorption then calculated via a thermo loop. We remove adsorbates where dispersion plays a large role (e.g. naphthalene, alkane, and CH$_3$I).

(4) \textbf{LC-RPA}: Ten surface species at the low surface (LC) coverage, calculated by RPA.\cite{schmidt2018benchmark}

Details of the data and preprocessing are shown in Supporting Information. 
 
\subsection{Representation}
In this study, we use a fixed length vector to encode the electronic and configuration information of the adsorbate and the metal facet. The adsorbate geometry is represented by the pathway fingerprint, a 2D graph-based representation that counts the frequency of selected subgraphs and thereby describes the connectivity information of the adsorbate. The geometry configuration of each surface species is obtained from reported DFT calculations, and converted into SMILES strings, then Rdkit\cite{rdkit} is used to count the occurrence of different subgraphs. In the pathway fingerprint, the list of subgraphs is constructed by enumerating all possible linear substructures up to path length of 3. The bond order information is neglected because it is hard to define the bond order between the adsorbate and metal without detailed bond length information which was not always reported. All hydrogen atoms in the substructures are explicitly accounted for. We use a generic identifier ([M]) to denote the metal atom in the SMILES strings without differentiating the element at this step. Then three integers are used: the group number (Co for 9, Ni for 10, Cu for 11, etc.), the period number (Cu for 4, Ag for 5, Au for 6, etc.), and the coordination number of the surface atoms to distinguish between the metal facets ((111) for 9, (100) for 8, (211) for 7, etc.) The surface coverage effect is encoded in one number, one over the square root of the surface coverage, which is equivalent to the number of atoms of the edge in the unit cell from DFT calculation (fractional coverage $\frac{1}{4}$ correspond to $2\times2$ unit cell, the number is 2). At the end, one integer denotes the number of layers in the quantum calculation, for experimental data, the number is set to be four, which is the most commonly used in DFT calculations. In summary, the feature is expressed as 

\begin{multline}
\label{eqn:repr}
    [\text{pathway fingerprint}, \\
    \text{group number},\text{period number}, \text{coordination number}, \\
    \sqrt{\frac{1}{\text{surface coverage}}}, \text{\# of layers}]
\end{multline}

\subsection{Training and Testing}
In following section, we evaluate the model performance on the task with high fidelity data, referred as target task. For ST-GP, the only task is the target task. For MT-GP, the task with high fidelity data is the target task, other tasks with low fidelity data are referred as source tasks. 

K-fold cross validation is used to calculate the generalization error of model prediction. The target task is randomly split into k parts and the source tasks do not change. Then, k-1 parts of target task and the whole source tasks are used to training the GP, and the left part of target task is predicted. The procedure is repeated k times. The model error is assessed through root mean squared error (RMSE), and mean absolute error (MAE) with cross validation prediction, such as 

\begin{align}
\label{eqn:rmse+mae}
    \text{RMSE} &= \sqrt{\frac{1}{N}\sum_{i=1}^{N}(y_{\text{pred}}^i-y_{\text{actual}}^i)^2}\\
    \text{MAE} &= \frac{1}{N} \sum_{i=1}^{N}|y_{\text{pred}}^i-y_{\text{actual}}^i|
\end{align}

The model performance in quantifying the model error is assessed by the mean of predicted standard deviation (MPSd)
\begin{equation}
    \text{MPSd} = \frac{1}{N} \sum_{i=1}^{N}\sigma(y_{\text{pred}}^i)
\end{equation}

To fully investigate the performance of MT-GP, We consider two cases studies. Case I is on purely computational dataset, FC. FC-RPA is the target task and FC-BEEF and FC-PBE serve as source tasks. Case II is on the combination of computational and experimental dataset. For ST-GP, PW91 and Exp is studied as target task. For MT-GP, Exp is the target, and PW91, LC-RPA are the source tasks. For Exp as target task, leave-one-out cross validation is used due to the limited size of the dataset. For the other case, 10-fold cross validation is used. When fitting the ST-GP and and MT-GP, 50 initial sampling points are used in minimizing the marginalzied likelihood. The rank of the coregionalized matrix is set to the number of tasks for the flexibility of the model.  

In this study, the variance (noise) of the output is treated as a hyperparemeter in the model, determined by optimization routine. For some cases, the noise could be known in prior, such as measurement error in experimental data. We should note that GP provides a flexible way to incorporate the heteroscedastic noise information.

\section{Results}

\subsection{Case I: Gaussian Process on FC-RPA}
\subsubsection{ST-GP on FC-RPA}
\begin{figure}
\includegraphics[scale=1.0]{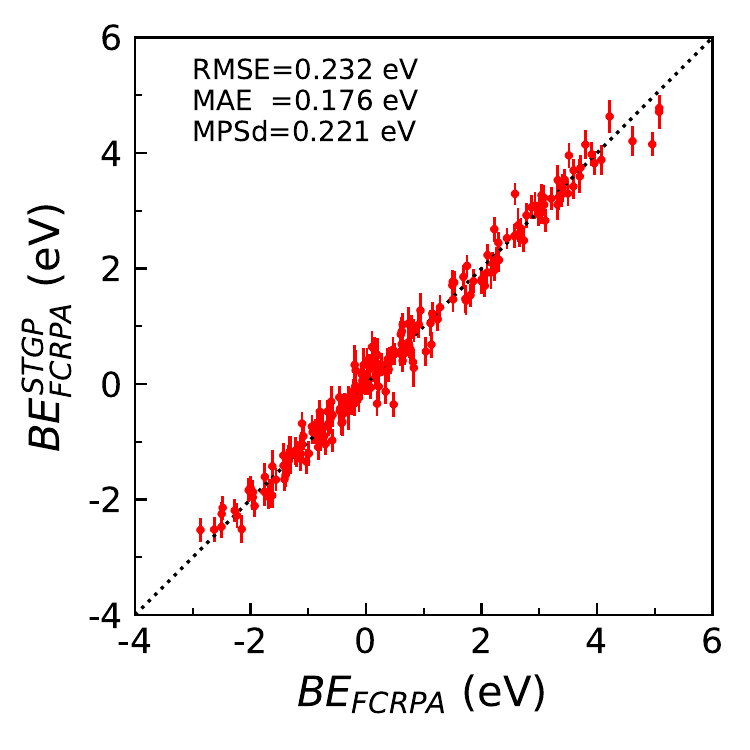}
\caption{Parity plot between FC-RPA dataset and ST-GP prediction evaluated by 10-fold cross validation. The prediction uncertainty corresponds to $\pm$ 1 std.}
\label{fig:st-rpa}
\end{figure}

To demonstrate the predictive power of Gaussian Process on adsorption energy, first we build the ST-GP on the FC-RPA dataset. FC is exhaustively generated with eight adsorbates on twenty-five transition metals, calculated by different levels of theory. Fig \ref{fig:st-rpa} demonstrates the prediction performance of ST-GP on FC-RPA. The prediction RMSE (0.23 eV) is close to MPSd (0.22 eV), which indicates that the predicted variance is able to capture the error of the model. This result demonstrates the promising performance of ST-GP on computational adsorption dataset.

\subsubsection{MT-GP on FC-RPA}

We further investigate the performance of multi-fidelity GP modeling on FC-RPA. We choose FC-PBE and FC-BEEF as low fidelity data in the separate task of MT-GP. PBE\cite{perdew1996generalized} belongs to a common GGA functional and is widely used in heterogeneous catalysis\cite{cheng2008bronsted, sabbe2012first, hammer1999improved, hammer2000theoretical, norskov2009towards}. BEEF-vdW is a GGA functional with van der Waals dispersion correction, which is fitted to a set of surface binding energies.\cite{wellendorff2012density} To investigate the influence on the choice of low level of theory, we build three MT-GPs with RPA as target task, with FC-PBE, FC-BEEF and the combination of FC-PBE and FC-BEEF as source tasks. 

\begin{figure*}
\includegraphics[scale=1.0]{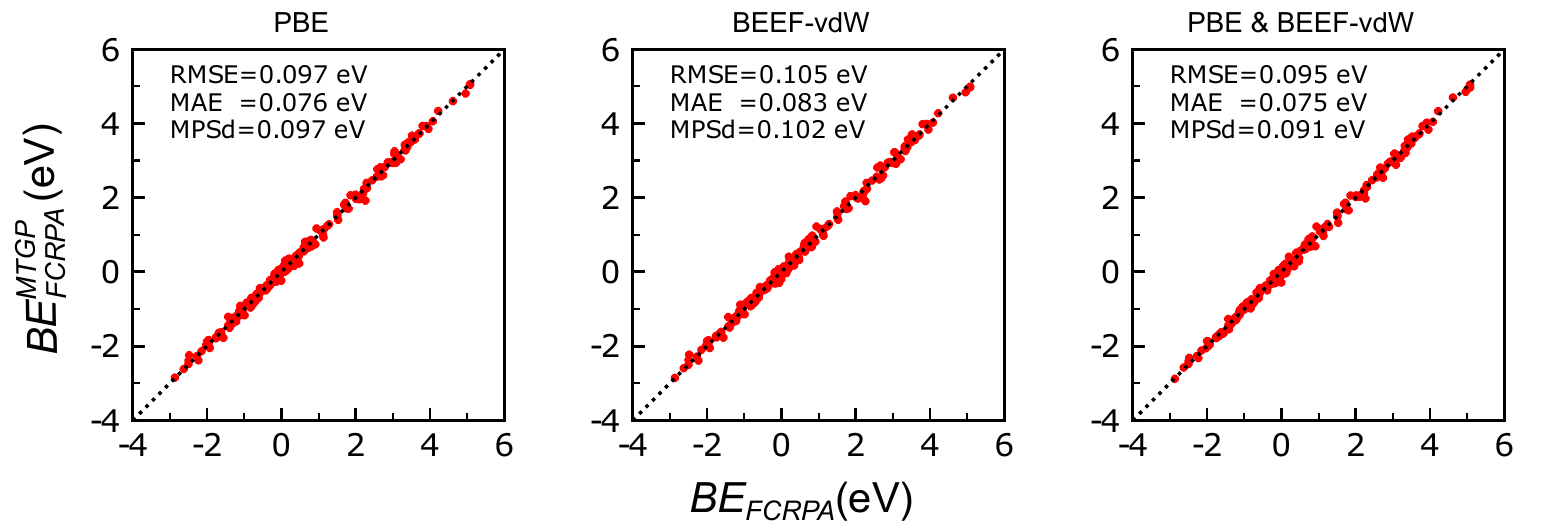}
\caption{Parity plot of between FC-RPA dataset and MT-GP model prediction with PBE (Left), BEEF-vdW (Center), and the combination of PBE and BEEF-vdW (Right), as source tasks. The prediction is evaluated with 10-fold cross validation}
\label{fig:rpa-subplot}
\end{figure*}

Fig. \ref{fig:rpa-subplot} shows the comparison of the three MT-GP models. The prediction error of all three MT-GPs are similar (~0.1 eV) and better than the ST-GP model discussed earlier. Combining FC-PBE and FC-BEEF as source tasks only marginally improves model performance over using them individually. We should note that increasing the number of tasks in MT-GP will increase the number of hyperparameters, which in turn significantly increases the computational cost for optimization and model training; this ultimately can lead to overfitting.

\begin{figure*}
\includegraphics[scale=1.0]{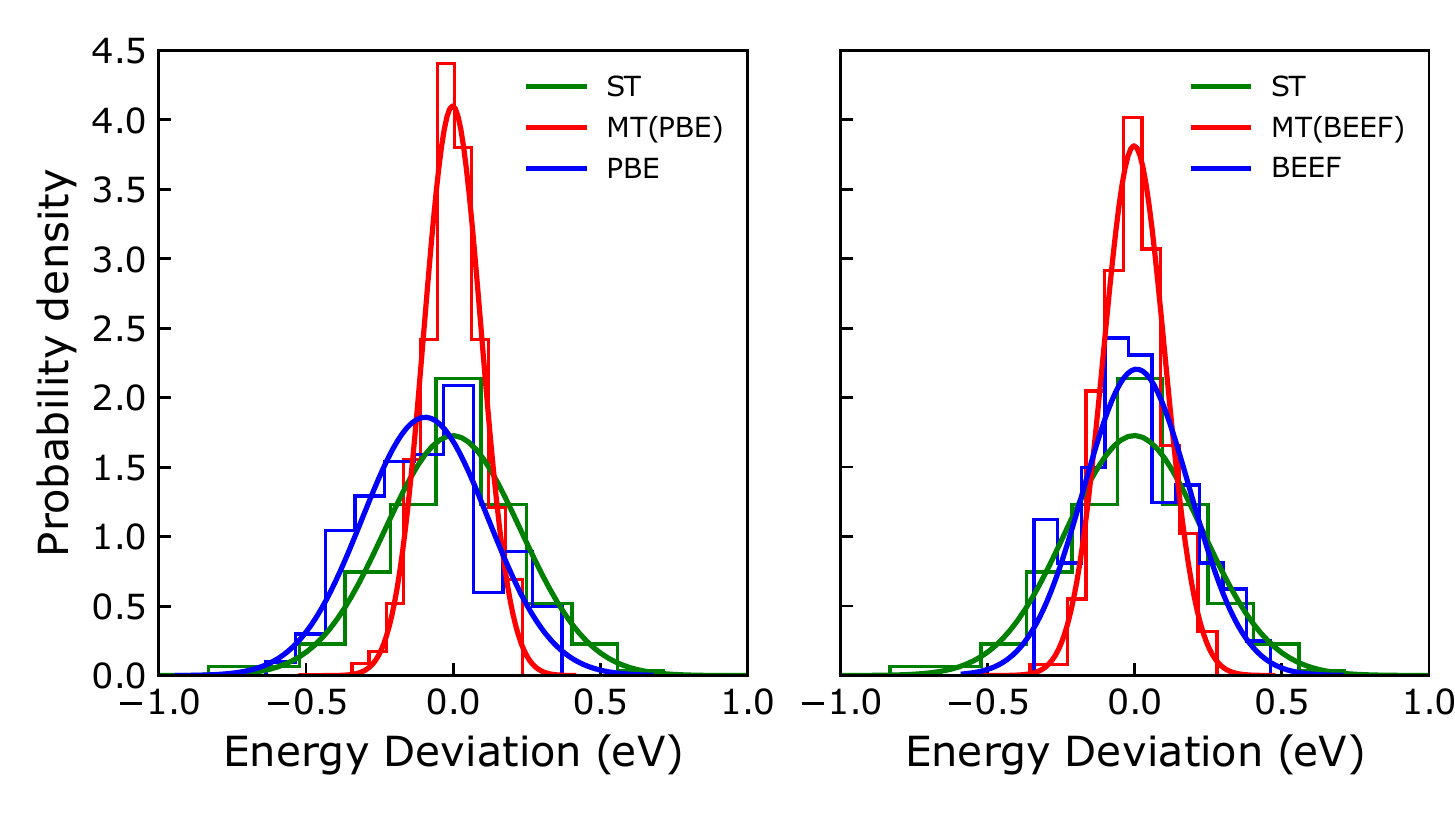}
\caption{Probability distribution of the deviations between RPA and (left) STGP, FC-PBE, and MTGP with FC-PBE as sources task; (right) STGP, FC-BEEF, and MTGP with FC-BEEF as sources task a. The histogram is fitted with normal distribution}
\label{fig:errdis-rpa-compare}
\end{figure*}

Fig. \ref{fig:errdis-rpa-compare} demonstrates the comparison of the error distributions of ST-GP, MT-GP and DFT functionals vs. FC-RPA. Clearly, we could see ST-GP and FC-PBE have similar prediction performance (0.23 vs. 0.21 eV in RMSE), but the bias of FC-PBE is larger than ST-GP (-0.09 vs. 0.00 eV), and FC-BEEF is slightly better than ST-GP (0.18 vs. 0.21 eV for RMSE, 0.1 vs 0.0 eV for bias). Both MT-GPs outperform the ST-GP and their source task, demonstrating that both high level and low level fidelity data contribute to the improvement of the MT-GP model prediction.

\subsection{Case II: Gaussian Process on PW91 and EXP}

Next, we investigate the performance of GP for ``heterogeneous data", viz., comprising PW91, EXP, and LC-RPA datasets.

\subsubsection{ST-GP on PW91}
\begin{figure}
\includegraphics[scale=1.0]{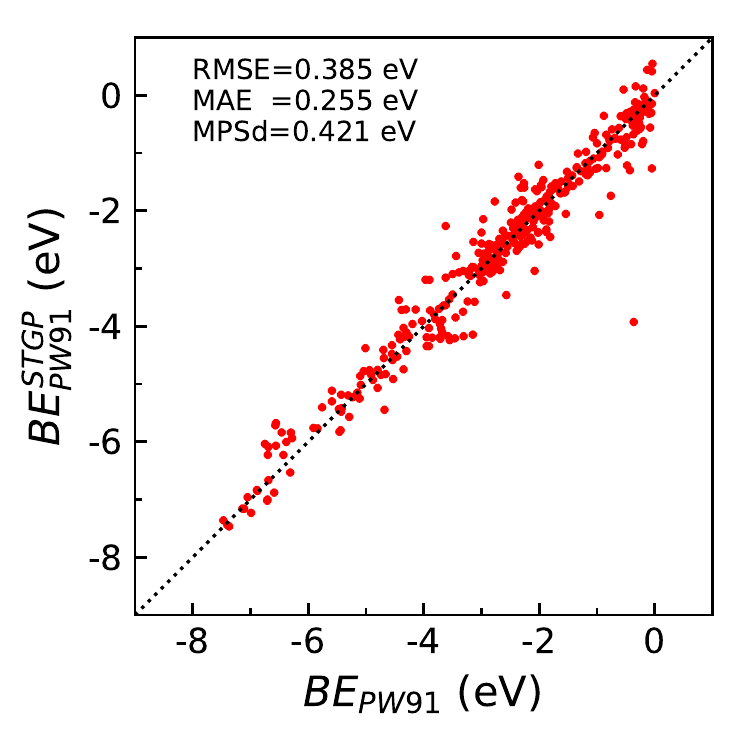}
\caption{Parity plot between PW91 dataset and ST-GP model prediction evaluated with 10-fold cross validation}
\label{fig:st-pw91}
\end{figure}

Fig. \ref{fig:st-pw91} shows that a ST-GP model can predict PW91 binding energies using our fairly simplistic fingerprints and SE covariance function with an average error of 0.26 eV. The average variance and lengthscale is 26.10 and 3.9497, and the maximum Euclidean distance between the molecules is 10.67. From Eqn. \ref{eqn:stgp_rbf}, the kernel element of the most distant pairs of species in the dataset is about 0.456, while the kernel element of the closest pairs is 681. The separation scale of the kernel matrix element indicates the diversity of the training set. 

\begin{figure}
\includegraphics[scale=1.0]{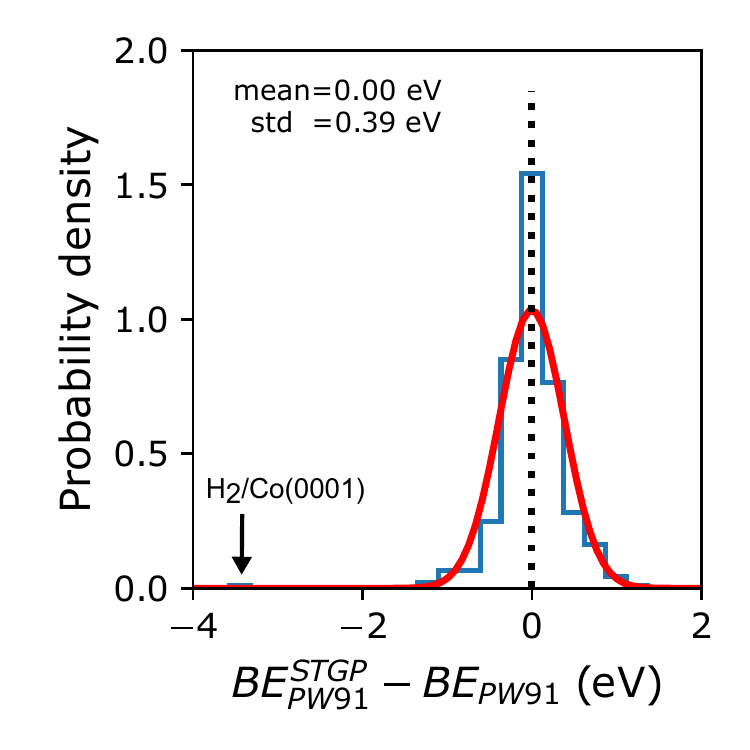}
\caption{Probability distribution of ST-GP model error of PW91 dataset. The histogram is fitted with normal distribution. The outlier is indicated with arrow.}
\label{fig:errdis-st-pw91}
\end{figure}

The prediction errors of the ST-GP model on PW91 are about 0.1 eV larger than ST-GP for the FC-RPA data, which is likely due to a larger diversity of the species considered in the PW91 set. For FC-RPA dataset, every adsorbate is obtained with monolayer surface coverage, the geometry configuration of the adsorbate is linear, which is relatively simple and easily captured by pathway fingerprints. And for PW91 dataset, the surface species is with low surface coverage, several adsorbate with nonlinear geometry have been studied, such as -COOH, -HCOO, etc. Further, the surface coverage and the coordination number has an equal representation in the FC-RPA dataset, but is not quite ensured in the PW91 set; this can affect the cross validation results in the case of random partitioning. For example, Fig. \ref{fig:errdis-st-pw91} shows the distribution of the error of ST-GP on PW91. We can clearly notice an outlier in the plot, corresponding to the prediction of H$_2$/Co(0001) with an absolute error 3.61 eV. The reported binding energy for H$_2$/Co(0001) is 0.36 eV, and predicted binding energy is 3.97 eV. The reason for the large error is that only one H$_2$ adsorbate appears in the dataset. The predictive standard deviation is 1.2 eV, which is underestimated. Including more data points pertaining to the molecular adsorption of H$_2$ on other transition metals can, in principle, resolve this discrepancy. We show in the supporting information that an ST-GP model trained on the PW91(e) dataset, which contains an additional data point involving molecular H$_2$ (on Fe(110) facet), can indeed fix this issue. Alternatively, more carefully designed representations with greater structural information could also resolve this;\cite{faber2017prediction} however, such an exercise is not the goal of this research.

\subsubsection{ST-GP on EXP}

\begin{figure}
\includegraphics[scale=1.0]{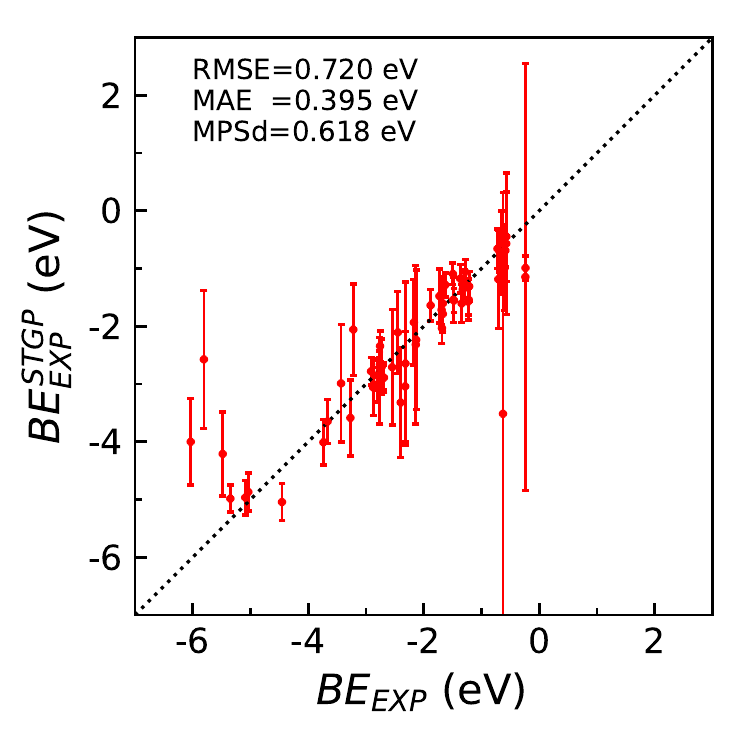}
\caption{Parity plot between Exp dataset and ST-GP model prediction evaluated with leave-one out cross validation. The prediction uncertainty corresponds to $\pm$ 1 std.}
\label{fig:st-exp}
\end{figure}

\begin{figure}
\includegraphics[scale=1.0]{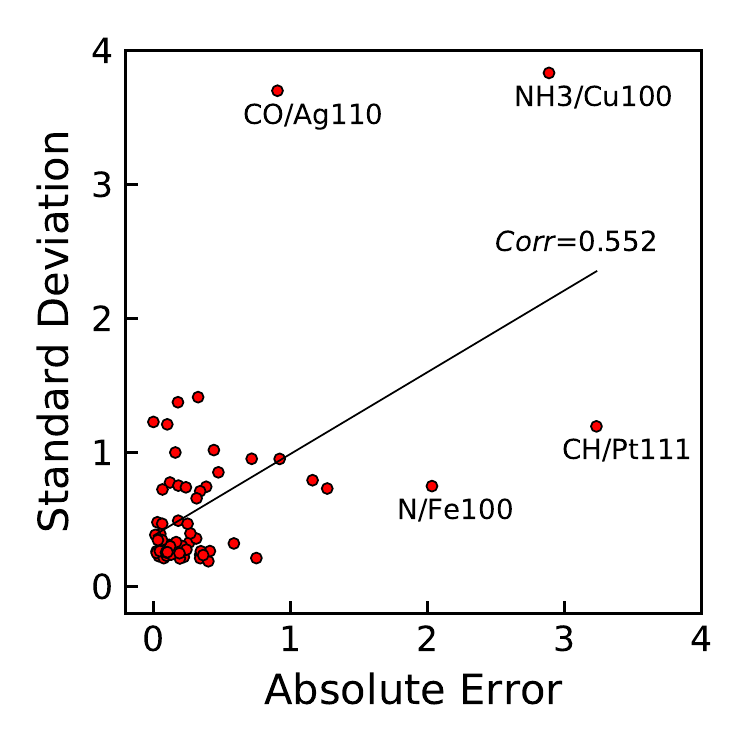}
\caption{Correlation between the absolute prediction error and predicted variance of ST-GP model on Exp dataset. The prediction uncertainty corresponds to $\pm$ 1 std.}
\label{fig:corr-st-exp}
\end{figure}

We applied our ST-GP formulation on the experimental adsorption database (Exp), wherein the data is relatively limited. Fig. \ref{fig:st-exp} shows that the error and variance of ST-GP prediction of the Exp dataset is much larger than both PW91 and FC-RPA dataset. One main reason is due to the size of the dataset, and the presence of outliers. Fig. \ref{fig:corr-st-exp} shows the correlation between the model prediction error and GP predicted standard deviation. The correlation between the error and predicted variance shows the GP model can identify the potential molecules with large error. We can see the model underestimates the binding energy for strong chemisorbed species, such as N/Fe(100) and CH/Pt(111), and overestimates for weakly binding species (NH$_3$/Cu(100)) due to insufficient data. Intriguingly, large prediction variance is seen for CO/Ag(111). This cannot be from the lack of similarity for geometry configuration, since fourteen other data points involve CO. We note that the surface coverage for this measurement is very low (nearly 0 coverage)\cite{silbaugh2016energies} relative to other data points, which inflates the corresponding entry in the representation vector of this species in the GP model. 

Next, we present multi-fidelity modeling to improve the model prediction by adding computational data.

\subsubsection{MT-GP on EXP}

Here, we perform multi-fidelity GP on experimental dataset with PW91 and LC-RPA as source tasks. In MT-GP on FC-RPA, the data in the source tasks cover the target task; that is, for every point in the source task, there is a corresponding point in the target task. In MT-GP on EXP, however, this is not the case. In this situation, each task corresponds to different set of inputs, usually referred as "asymmetric" multitask learning\cite{alvarez2012kernels, xue2007multi}. As noted in the previous section, MT-GP is a flexible framework which allows to predict binding energy with high-level accuracy at points for which low fidelity data is not available.

\begin{figure}
\includegraphics[scale=1.0]{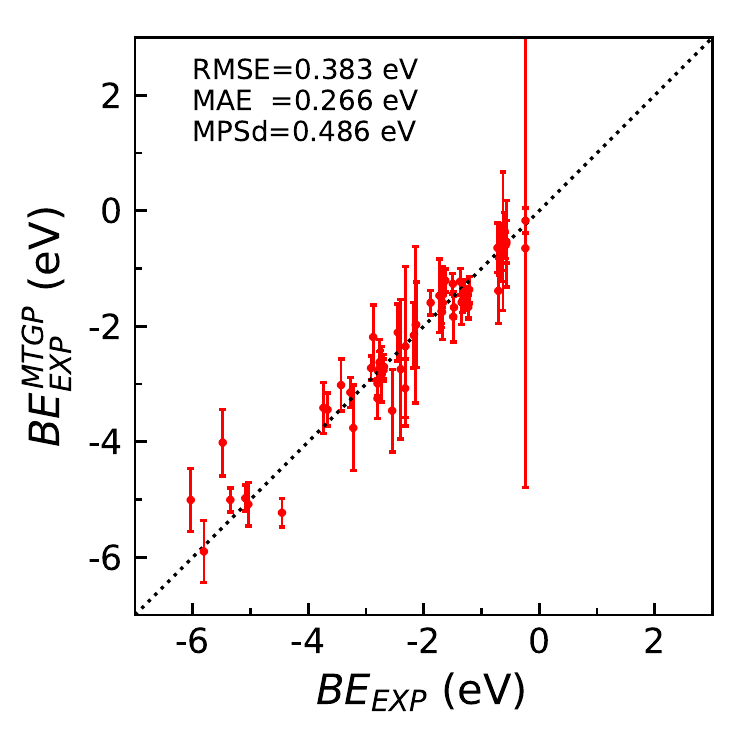}
\caption{Parity plot between Exp dataset and MT-GP model prediction with PW91 as source task. The performance is evaluated with leave-one-out cross validation}
\label{fig:mt-exp}
\end{figure}

Fig. \ref{fig:mt-exp} demonstrate the MT-GP prediction on Exp dataset with PW91 as source task. The error and variance of the MT-GP model are significantly lower than those of the single task models. The error of the outliers in ST-GP have been reduced. However, we could see that some species with large variance still remain, such as CO/Ag(111); the reason is very simple in that some features in those molecules does not cover in the PW91 dataset. In such a case, the representation may not enough to capture the extreme low surface coverage of the species. To circumvent this, bringing more electronic features into the representation is one possible solution. 

To test the influence of the large variance data and outliers, we do the statistical analysis on the prediction data with standard deviation less than 0.5 eV, finally 43 data points left, with RMSE about 0.25 eV. After filtering the data with large variance, only O/Rh(100) has absolute error larger than 0.5 eV, which is not shown in the PW91 dataset. There are 33 species in both EXP and PW91 dataset, the RMSE is 0.37 eV, suggesting MT-GP improves the prediction over the source tasks.

Then, we perform the MT-GP with the expanded PW91 dataset and LC-RPA as source task. PW91(e) dataset contains 131 more data points than PW91 dataset, which carries more information, but adding noise by including the species calculated with different settings (e.g. Monkhorst Pack k-points sampling). In addition to FC-RPA dataset, Schmidt also perform RPA calculation on ten surface species with experimentally relevant (low surface) coverage, with MAE 0.21 eV compared to experimental dataset.\cite{schmidt2018benchmark} From Tbl. \ref{tbl:summary}, we could see the model error using PW91(e) and LC-RPA as source task has been reduced further from PW91 as source task. Interestingly, the MPSd follows the same trend as including more source tasks; the GP models, therefore, provide reliable uncertainty estimates. It is quite remarkable that the multi-task GP is able to learn from diverse data sources. As the size of training data increases (even if acquired from different sources), the model performance can constantly improve.

\begin{table*}[!ht]
\begin{tabular}{|c|c|c|c|c|c|c|}
\hline
Target Task                & Model               & Source Task    & RMSE (eV)   & MAE (ev)   & MPSd (eV) & lengthscale   \\ \hline
\multirow{4}{*}{FC-RPA}    & ST                  & --             & 0.232       & 0.176      & 0.221     & 4.204         \\ \cline{2-7} 
                           & \multirow{3}{*}{MT} & PBE            & 0.097       & 0.076      & 0.097     & 2.151         \\ \cline{3-7} 
                           &                     & BEEF           & 0.105       & 0.083      & 0.102     & 2.165         \\ \cline{3-7} 
                           &                     & PBE \& BEEF    & 0.095       & 0.075      & 0.091     & 1.651         \\ \hline 
\multirow{5}{*}{Exp}       & ST                  & --             & 0.720       & 0.395      & 0.618     & 5.166         \\ \cline{2-7} 
                           & \multirow{4}{*}{MT} & PW91           & 0.383       & 0.266      & 0.486     & 3.846         \\ \cline{3-7}
                           &                     & PW91(e)        & 0.329       & 0.245      & 0.448     & 3.776         \\ \cline{3-7} 
                           &                     & PW91 \& LC-RPA & 0.380       & 0.257      & 0.471     & 3.704         \\ \cline{3-7} 
                           &                     & PW91(e) \& LC-RPA & 0.321    & 0.232      & 0.437     & 3.688         \\ \hline 
PW91                       & ST                  & --             & 0.418       & 0.272      & 0.400     & 4.006         \\ \hline
PW91(e)                    & ST                  & --             & 0.355       & 0.252      & 0.408     & 3.973         \\ \hline

\end{tabular}
\caption{Summary of the model prediction}
\label{tbl:summary}
\end{table*}

\section{Discussion}

\subsection{Learning from Homogeneous and Heterogeneous dataset}
Clearly, Gaussian Process models allow us to predict the binding energy from different level of fidelity data of GGA functional(PW91), high level of functional (RPA), and experimental data, using pathway fingerprints and electronic feature of transition metal facet as the representation. The prediction performance depends on the size and diversity of the dataset. For ST-GP on FC-RPA dataset, the error and variance values are smaller than that of ST-GP on PW91 and experimental datasets. Due to the diversity and the incompleteness of the latter sets, the sampling bias is large, and the prediction variance is large and not evenly distributed. This situation suggests the generation of a exhaustive dataset for surface species which evaluated at different level of theory, such as PBE, PW91, BEEF-vdW and RPA.

\subsection{Learning from the correlation between tasks}

\begin{figure}
\includegraphics[scale=1.0]{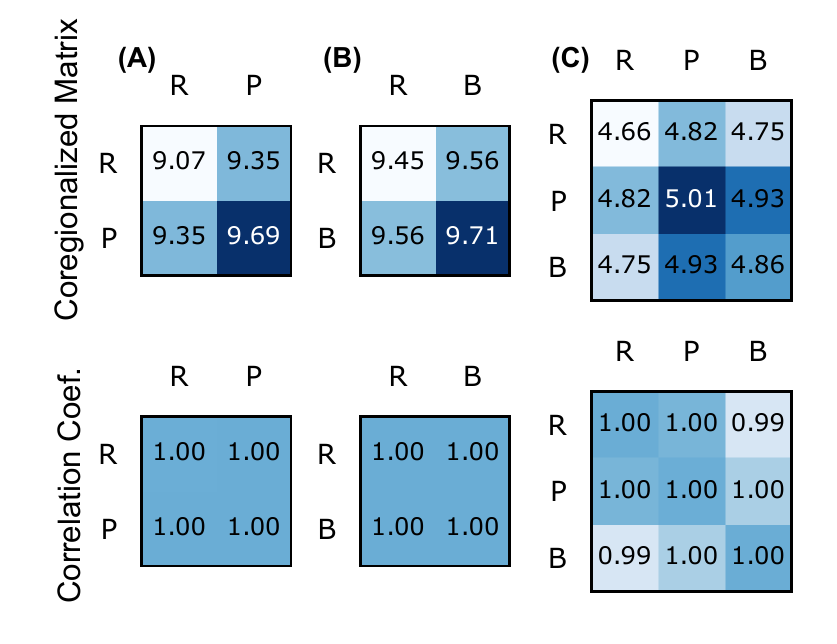}
\caption{Correlation structure in MT-GPs with different low level of theory. Source task: (A) FC-PBE; (B) FC-BEEF; (C) FC-PBE and FC-BEEF. Task abbreviation: R (FC-RPA); P (FC-PBE); B (FC-BEEF).}
\label{fig:rpa-cov}
\end{figure}

One key factor of MT-GP is that it learns the correlation and similarity among each tasks. When making a prediction, MT-GP weights different tasks according to coregionalized matrix, and all data points (across all tasks) will contribute to the prediction for the target task. Here, we illustrate the MT-GP performance on FC-RPA dataset. Fig. \ref{fig:rpa-cov} shows the coregionalized matrix $\mathbf{B}$ in Eqn. \ref{eqn:mtgp_K} and the normalized correlation matrix. The off-diagonal element of correlation matrix is very close to 1, meaning that the GP model captures the strong correlation and similarity between the source and target tasks. In the original dataset, the Pearson coefficient is 0.9948 between FC-PBE and FC-RPA, and 0.9953 between FC-BEEF and FC-RPA. The off-diagonal element is slightly higher than the original Pearson coefficient because the MT-GP model assigns part of the difference between different tasks as random noise to increase the generalization of the prediction.
 
Fig. \ref{fig:rpa-subplot} and \ref{fig:errdis-rpa-compare} has shown the comparison of the performance of two MT-GPs with FC-PBE and FC-BEEF as source task. We can see the model prediction error using FC-PBE as source task is slightly smaller than using FC-BEEF as source task. We do not expect this result, as shown in Fig. \ref{fig:errdis-rpa-compare}, since the mean of the difference between FC-PBE and FC-RPA (0.09 eV) is smaller than 0.00 eV between FC-BEEF and FC-RPA. Also, the standard deviation is smaller for FC-BEEF (0.18 eV) than FC-PBE (0.21 eV). One reason we hypothesize is that BEEF-vdW is specifically designed for surfaces having been parameterized to surface science dataset. The systematic error of BEEF-vdW against surface catalysis dataset is reduced by adjusting the coefficients in the functional. Besides, PBE and RPA are both free-parameter functionals with \textit{ab initio} nature. As not captured by the correlation of MT-GP with single functional in source task, the normalized terms in both MT-GP are 1.0. But shown in the MT-GP with both FC-PBE and FC-BEEF as source tasks, the correlation in MT-GP between FC-PBE and FC-RPA is stronger than the term between FC-BEEF and FC-RPA.

\subsubsection{Learning with few data points}

Pilania has shown that the MT-GP model performance will progressively improve as increasing the number of the training data in both target and source tasks.\cite{pilania2017multi} We are interested in designing the training dataset with less but informative data points with high fidelity to reach similar performance as using the full dataset. 

Here we adapt the active learning framework, also known as optimal experimental design, using the predictive variance as acquisition criterion to iteratively build the training dataset. We perform active learning on MT-GP with FC-RPA as target task, FC-PBE as source task. To do that, we randomly pick 50, 100, 150, and 200 FC-PBE data points as source task, then iteratively add the data of FC-RPA level with largest predictive variance into the training set, and measure the prediction error on the remaining FC-RPA dataset. To avoid sampling bias, the procedure is repeated 5 times by resampling the initial FC-PBE dataset.

\begin{figure*}
\includegraphics[scale=1.0]{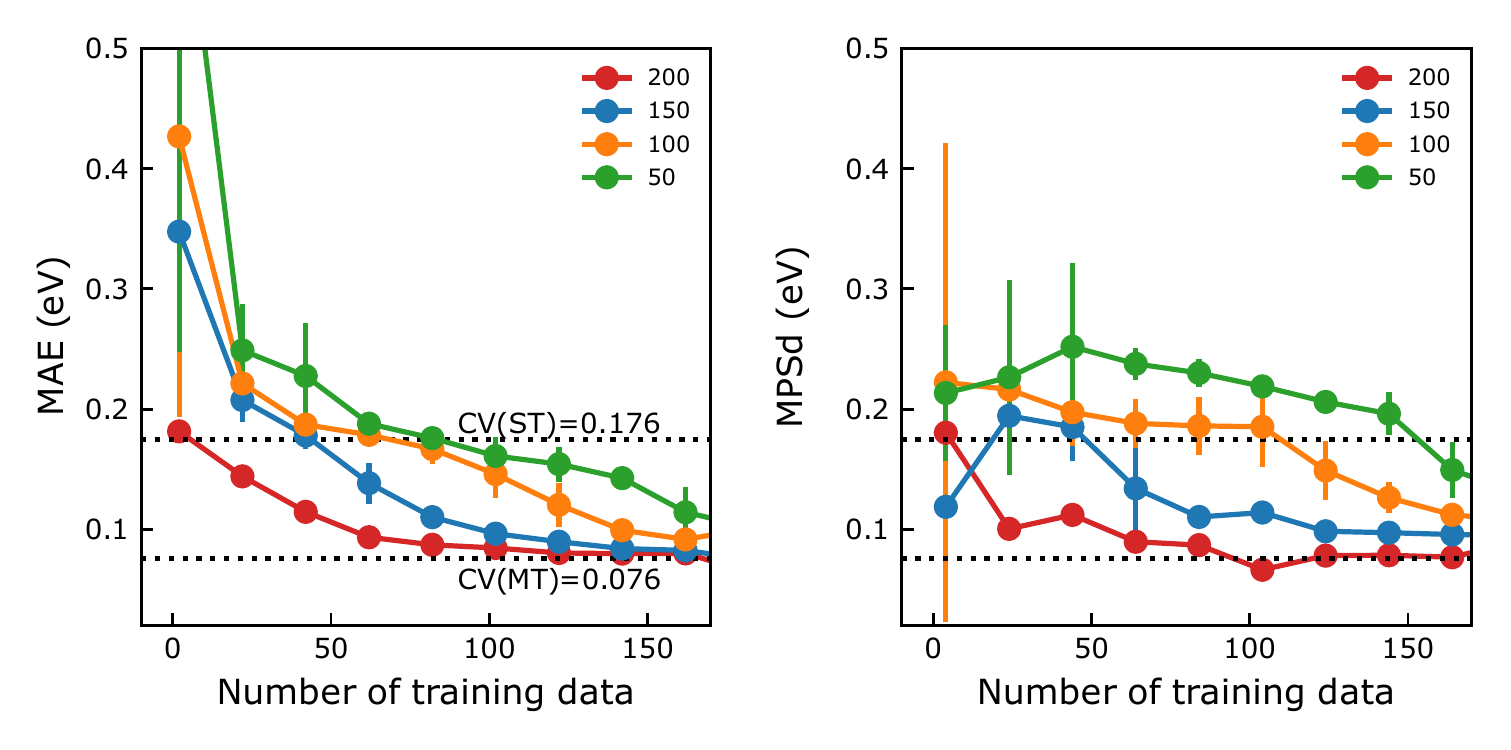}
\caption{MAE (left) and MPSd (right) on the remaining FC-RPA dataset along active learning curve with different number of FC-PBE data as source task}
\label{fig:rpa-active}
\end{figure*}

Fig. \ref{fig:rpa-active} demonstrates the learning performance with different number of FC-PBE data points in the source task. With full (200) PBE dataset as source task, MT-GP is able to reach the same level of accuracy as ST-GP using less than ten FC-RPA data points. With about sixty FC-RPA data points, the model error is close to the 10-fold cross validation error of MT-GP using 180 FC-RPA, full FC-PBE. This clearly demonstrates the value of active learning in designing the training data. The mean predicted standard deviation (MPSd) is smaller than the model error during the early stages of learning, which indicates the overfitting problem when the data is scarce. Subsequently, MPSd captures MAE fairly well; MPSd is, therefore, a good indicator of when to stop the learning process. In MAE and MPSd, as we expect, the learning curve shifts upside when decreasing the number of PBE in source task. Comparing different learning curves, we could see that the number of data point in low level theory is an important determinant of the performance of the multi-fidelity model. To reach the same level of accuracy (say 0.12 eV), 60, 100, and 140 FC-RPA data points are needed if the source task has 200, 150, and 100 FC-PBE points respectively. As shown here, MT-GP could reach high level accuracy with fewer, but carefully selected high fidelity data by fusing low fidelity information. Also, as increasing the number of the data with low fidelity, the number of data with high fidelity needed reduced to reach the same level of accuracy.

\section{Conclusion}
In this paper, we have exploited the multi-fidelity modeling on adsorption energy database. Pathway fingerprint together with electronic information of metal facet has been employed to represent the surface species. Gaussian Process provides a flexible probabilistic framework to incorporate different level of density functional theory and accurate experimental measurement. We test the MT-GP on two dataset, (1) exhaustively generated dataset calculated by RPA, a high level of theory which recently shown promising on adsorption. (2) experimental dataset and PW91 function collected from several computational benchmark papers. In both cases, the multi-task Gaussian Process model performed substantially better than any single-task models (trained on one set of data). Further, it was shown that progressively fewer high fidelity data points was required to reach a specified level of accuracy as more low fidelity data was included in the training. Multi-task Gaussian Processes, therefore, offer an effective means of training data-driven models using a combination of data from various sources and comprising of different fidelity. 

\section{Conflict of Interest}
The authors declare no conflict of interest.

\section{Acknowledgement}
This work was supported by the startup package from Lehigh University.

\bibliography{MT-GP}
\end{document}